\documentclass[12pt]{iopart}
\usepackage{graphicx}
\usepackage{cite}

\begin{document}

\title[Relativistically transparent magnetic filaments]{Relativistically transparent magnetic filaments: scaling laws, initial results and prospects for strong-field QED studies}

\author{H. G. Rinderknecht$^1$, T. Wang$^2$, A. Laso Garcia$^3$, G. Bruhaug$^1$, M. S. Wei$^1$, H. J. Quevedo$^4$, T. Ditmire$^4$, J. Williams$^5$, A. Haid$^5$, D. Doria$^6$, K. Spohr,$^6$ T. Toncian$^3$ and A. Arefiev$^2$}

\address{$^1$University of Rochester Laboratory for Laser Energetics, Rochester, NY 14623}
\address{$^2$University of California, San Diego}
\address{$^3$Helmholtz-Zentrum Dresden-Rossendorf, Germany}
\address{$^4$University of Texas, Austin}
\address{$^5$General Atomics, San Diego}
\address{$^6$ELI-NP, Romania}

\begin{abstract}
	Relativistic transparency enables volumetric laser interaction with overdense plasmas and direct laser acceleration of electrons to relativistic velocities.
	The dense electron current generates a magnetic filament with field strength of the order of the laser amplitude ($>$10$^5$~T).
	The magnetic filament traps the electrons radially, enabling efficient acceleration and conversion of laser energy into MeV photons by electron oscillations in the filament.
	The use of microstructured targets stabilizes the hosing instabilities associated with relativistically transparent interactions, resulting in robust and repeatable production of this phenomenon.
	Analytical scaling laws are derived to describe the radiated photon spectrum and energy from the magnetic filament phenomenon in terms of the laser intensity, focal radius, pulse duration, and the plasma density.
	These scaling laws are compared to 3-D particle-in-cell (PIC) simulations, demonstrating agreement over two regimes of focal radius.
	Preliminary experiments to study this phenomenon at moderate intensity ($a_0 \sim 30$) were performed on the Texas Petawatt Laser.
	Experimental signatures of the magnetic filament phenomenon are observed in the electron and photon spectra recorded in a subset of these experiments that is consistent with the experimental design, analytical scaling and 3-D PIC simulations.
	Implications for future experimental campaigns are discussed.  
	
\end{abstract}

\vspace{2pc}

\vspace{1pc}
\noindent{\it Version:} \today

\submitto{\NJP}

\maketitle

\section{Introduction}
\label{sec:Intro}
High-power lasers enable a new regime of relativistically transparent laser--plasma interaction physics. 
Collective electron motion typically prevents the propagation of electromagnetic waves in a plasma if the frequency is below the electron plasma frequency, $\omega_{\mathrm{pe}} = \sqrt{n_{\mathrm{e}} e^2/\epsilon_0 m}$.
This limitation sets a critical plasma density ($n_{\mathrm{c}}$) that a low-intensity laser cannot penetrate, which depends on the laser wavelength $\lambda$ as: $n_{\mathrm{c}} [\textrm{cm}^{-3}] \approx 1.1\times10^{21} / (\lambda [{\mu}\textrm{m}])^2$.  
However, if the laser electric field $E$ is sufficiently strong that the normalized laser amplitude $a_0 = eE/\omega m c > 1$, electron motion within the laser field becomes relativistic and the critical density is increased due to the increase in the effective electron mass\cite{POP:Arefiev:2020}.
As such, the plasma remains transparent for density $n_{\mathrm{e}} < \gamma n_{\mathrm{c}}$, where the Lorentz factor $\gamma = \sqrt{1+a_0^2} \approx a_0$.
For optical lasers ($\lambda \sim 1~\mu$m), this relativistically transparent interaction requires intensity exceeding $10^{18}$~W/cm$^2$.

\begin{figure}
	\centering
	\includegraphics[trim=140 339 140 335,clip,width=\textwidth]{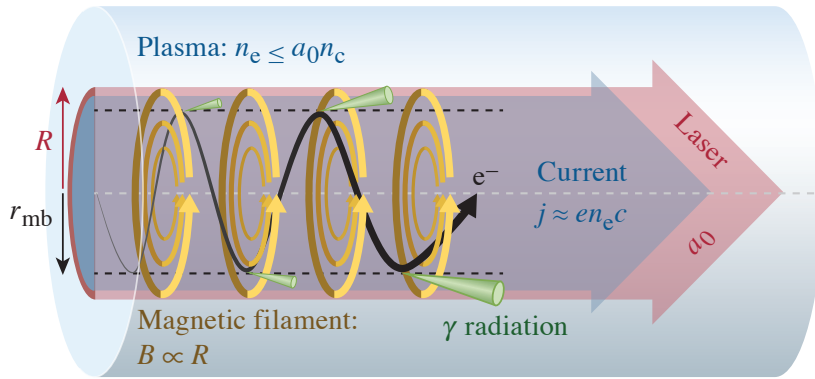}
	\caption{Schematic diagram of the relativistically transparent magnetic filament phenomenon.  An intense laser (red) drives a relativistic current (blue) in a relativistically transparent plasma.  The azimuthal magnetic field (orange) traps electrons (black) either within the laser radius $R$ or a magnetic boundary $r_{\mathrm{mb}}$ (see text).  Electrons are directly accelerated by the laser fields and radiate high-energy photons (green).\label{fig:cartoon}}
\end{figure}

Relativistic electron motion and volumetric laser--plasma interaction at super-critical densities combine to create a fascinating phenomenon: the relativistically transparent magnetic filament.\cite{POP:Arefiev:2016, PRL:Stark:2016, PRE:Huang:2016, PPCF:Jansen:2018, POP:Wang:2019, SR:Gong:2019, NJP:Rosmej:2019, POP:Wang:2020, PRE:Gong:2020, PRE:Arefiev:2020, PRA:Wang:2020}
A schematic diagram of this phenomenon is shown in Fig.~\ref{fig:cartoon}.
As the intense laser propagates through an overdense plasma, the ponderomotive force pushes the electrons forward, inducing a relativistic current filament that moves axially with the laser field.
This current in turn generates a quasi-static azimuthal magnetic field with strength comparable to the oscillating laser field.  
Surrounding the relativistically transparent plasma with a higher-density channel wall prevents the hosing instability and optically guides the laser pulse, allowing intense laser--plasma interaction over many Rayleigh lengths.
This system can be understood as a relativistic plasma rectifier for laser light that efficiently converts the oscillating electric and magnetic fields into a direct filamentary current and associated magnetic field.  
Electrons oscillate within this confining azimuthal magnetic field, facilitating the direct energy gain from the laser to ultrarelativistic energies (hundreds of MeV).
At the same time, the electrons' deflections within the magnetic field cause them to emit photons.
State-of-the-art simulations show that the azimuthal field strength reaches the mega-Tesla level and the effective acceleration gradient exceeds $10^5$~MeV/cm in the multipetawatt regime.
The extreme magnetic-field strength and high electron energy combine to boost the radiated photon energy $\langle \epsilon_\gamma \rangle \propto B \gamma^2$ and the radiative power $P \propto B^2 \gamma^2$, such that the magnetic filaments become efficient radiators of MeV photons.

Relativistically transparent magnetic filaments promise to provide both a highly efficient laser-driven source of gamma rays and a unique and robust platform for the study of high-field physics.
In 3D simulations, the conversion efficiency from laser energy into collimated MeV photons can exceed 1\%\cite{PRA:Wang:2020}.
This is substantially greater than presently demonstrated laser-driven sources of betatron and Compton x rays in the 100-keV to multi-MeV range, which typically feature efficiency of the order of $10^{-6}$ or less\cite{PPCF:Albert:2016}.
The prospect of generating copious multi-MeV photons in laser--target interactions has sparked particular interest due to its many potential scientific and technology applications, such as Breit--Wheeler pair production\cite{PRA:Wang:2020}, laboratory astrophysics\cite{PPR:Bulanov:2015}, photon--photon scattering\cite{PTEP:Homma:2016}, and photonuclear spectroscopy\cite{PRC:Schreiber:2000,PRC:Kwan:2011}, as well as medical applications\cite{MP:Weeks:1997, PMB:Girolami:1996}.
The physical state of the magnetic filaments is also remarkable in that they are capable of producing quasistatic magnetic fields of the order of $10^6$~T, comparable to the atmospheres of neutron stars\cite{RPP:Harding:2006}.
In the reference frame of the relativistic electrons, these extreme magnetic fields are further boosted by a factor of $\gamma$ such that they may approach or exceed the Schwinger critical field strength, $B_{\mathrm{c}} \approx 4.4\times10^9$~T.
In this quantum-dominated regime, radiation reaction becomes significant and a single photon can carry an appreciable fraction of the emitting electron's energy, dramatically altering particle-field interaction dynamics\cite{RMP:DiPiazza:2012}.
This platform therefore offers the potential to study the effects of strong-field quantum-electrodynamic (QED) physics on collective processes in extreme plasma conditions relevant to astrophysics.

While prior theoretical work on this phenomenon has typically focused on ultra-intense laser impulses that will soon be available ($5\times10^{22}$~W/cm$^2$, $a_0 = 190$), the phenomenon itself is robust for relativistically transparent interactions and is expected to scale from much lower laser intensity $a_0 \sim 10$.  
In this manuscript, we present an analytical scaling for the characteristic radiation produced by this phenomenon as a function of laser intensity and other design parameters.
We also present the results of recent experiments performed at moderate intensity to test this scaling, and discuss the implications of the scaling for future experimental work.
This manuscript is organized as follows:
Section~\ref{sec:Scaling} presents the derivation of analytical scaling laws for radiation from magnetic filaments, as well as the comparison of these laws with results from 3-D particle-in-cell (PIC) simulations.
Section~\ref{sec:Experiment} describes the design and initial results of  moderate-intensity experiments performed on the Texas Petawatt Laser (TPW).
Section~\ref{sec:Discussion} interprets the experimental results in the context of the derived scaling laws, and discusses the implications of this work for experiments at higher intensity.
Finally, Sec.~\ref{sec:Conclusions} concludes with a discussion of the prospects for application of this phenomenon for experimental strong-field QED studies.

\section{Scaling Model}
\label{sec:Scaling}
Intense laser fields can rapidly drive relativistic electron current filaments in plasmas, producing strong azimuthal magnetic fields.
The maximum current density $j$ that can be sustained in a plasma is limited by the electron density $n_{\mathrm{e}}$ as: $j < j_{\mathrm{max}} = e n_{\mathrm{e}} c$, due to the fact that the electron velocity cannot exceed the speed of light $c$.  
The use of relativistically intense lasers therefore increases the limiting current density by allowing interaction within plasmas with $n_{\mathrm{e}} < a_0 n_{\mathrm{c}}$, and ultrastrong magnetic fields may be produced. 
It is convenient to use the normalized current density $\alpha \equiv j / j_{\mathrm{c}}$, where $j_{\mathrm{c}}$ is the current density limit at the critical electron density, $j_{\mathrm{c}} = e n_{\mathrm{c}} c$.
The maximum normalized current density then follows as $\alpha \leq n_{\mathrm{e}} / n_{\mathrm{c}}$, which for relativistically transparent interactions may approach a limiting value of $a_0$.\footnote{This definition of $\alpha$ is reduced by a constant factor of $1/\pi^2$ compared to the definitions used in Refs.~\cite{POP:Wang:2020,PRE:Gong:2020}.}

The azimuthal magnetic-field strength at a radius $r$ from the center of a uniform current filament with total radius $R$ is calculated using Ampere's law as:  $B = \mu_0 j r/2$.
Using the above definitions, the maximum strength of a filamentary field relative to the laser field strength $B_0$ is
\begin{eqnarray}
	\frac{B}{B_0} = \frac{\pi R \alpha}{\lambda a_0}.
	\label{eq:B_norm}
\end{eqnarray} 
This formulation shows the importance of relativistic transparency, which enables a high azimuthal magnetic field to be produced on the same order of magnitude as the laser field.  
In this context, the ratio $(\alpha / a_0)$ acts as a similarity parameter when scaling the magnetic filament system with laser intensity.
Notably, this parameter is identical to the similarity parameter derived by Pukhov and Gordienko for ultrarelativistic laser--plasma interactions ($S=n_{\mathrm{e}}/n_{\mathrm{c}}a_0$) in the limit $(v_e / c) \equiv \beta \rightarrow 1$\cite{PTRS:Pukhov:2006}.
We adopt the notation $S_\alpha$ and refer to this as the relativistic transparency parameter.

A study of accelerating electron orbits in such a magnetic filament established that efficient electron acceleration requires the channel to have a current density that satisfies $S_\alpha > 0.01$.\cite{PRE:Gong:2020}
Under these conditions, the electrons observe net accelerating phases of the oscillating laser field for a large fraction of initial transverse momenta.
The maximum electron energy gain observed under these conditions was approximately $\gamma_{\mathrm{max}} \sim 10 a_0$ per half bounce.
The number of bounces experienced depends on the current density and the length of the channel: one half bounce occurs over an axial distance of roughly 10$\lambda$, and therefore over a duration of roughly $10\lambda/c$.
However, the energy gain as a function of initial electron momentum was observed to depend chaotically on the initial electron momentum after as few as two bounces.
If this behavior is similar to other chaotic acceleration processes, such as those that occur in solar coronae, it will produce a spectrum of electrons up to some cutoff energy that depends on the duration of the acceleration process.  
We represent the electrons as a thermal distribution $f(\epsilon_{\mathrm{e}}) = (n_{\mathrm{e}} / T_{\mathrm{e}}) \exp{\left(-\epsilon_{\mathrm{e}}/T_{\mathrm{e}}\right)}$ with characteristic energy $T_{\mathrm{e}}$ that changes with time.
Given that the characteristic unit of direct laser acceleration is $a_0$ and acceleration can at best proceed linearly with the number of bounces in the channel, $T_{\mathrm{e}} = (mc^2) C_T a_0 t_\nu$ is taken as a plausible hypothesis, with a scalar constant $C_T$ in the initial stages of the acceleration and the normalized interaction time $t_\nu = (ct/\lambda)$ in units of the laser period.
Simulations performed as part of the present work (see Sec.~\ref{sec:Experiment}) suggest a value $C_T \approx 0.08$.
This simple model of the electron acceleration provides a basis for the derivation of scaling laws for the emitted radiation.

\subsection{Derivation of radiation scaling laws}
The oscillation of the accelerating electrons in the strong azimuthal magnetic fields produces electromagnetic radiation.
This radiation may be characterized using similar expressions to those of synchrotron emission in a uniform magnetic field.
The key parameter is the normalized electron acceleration, $\chi = \gamma B/B_{\mathrm{c}} \equiv \gamma B'$.  
The radiated power spectrum has a shape characterized by the formula

\begin{eqnarray}
	\frac{\mathrm{d}P}{\mathrm{d}\epsilon_*} = \frac{4}{9}\alpha_{\mathrm{fsc}}\frac{mc^2}{\hbar} B' F\left(\frac{\epsilon_*}{\epsilon_{\mathrm{c}}}\right), 
	\label{eq:dPde_synch} \\
	F(x) \equiv \frac{9\sqrt{3}}{8\pi}x\int_{x}^{\infty}K_{5/3}(z)\mathrm{d}z,
	\label{eq:S_synch} \\
	\epsilon_{\mathrm{c}} \equiv \frac{3}{2} \chi \gamma m c^2 ,
	\label{eq:omega_c_synch} 	
\end{eqnarray}
where $\alpha_{\mathrm{fsc}}$ is the fine structure constant, and $F$ contains all information on the spectral shape [$\int F(x) \mathrm{d}x = 1$] and is peaked near the critical frequency $\epsilon_{\mathrm{c}}$, with $\epsilon_{\mathrm{peak}} \approx 0.29 \epsilon_{\mathrm{c}}$.

The maximum magnetic field observed by the electrons will be limited by their available orbits inside the magnetic filament.
Analytical models show the electrons are confined within a magnetic boundary, $ r_{\mathrm{mb}}/\lambda \approx \sqrt{\gamma_i/\alpha}/\pi$ for an initial electron momentum prior to acceleration $\gamma_i$\cite{PRE:Gong:2020}.\footnote{Here we neglect the effects of phase velocity differing from $c$, which can increase the magnetic boundary as the electron gains energy.}
This average initial momentum arises from the electrons' initial interaction with the laser pulse and will therefore scale as $\gamma_i = f_i a_0$ for a constant factor $f_i \sim 1$.
The smaller of $R$ or $r_{\mathrm{mb}}$ will determine the maximum magnetic field observed by the electrons. 
Using Eq.~(\ref{eq:B_norm}), the ratio of the magnetic field to the critical field can be determined in terms of the design parameters as

\begin{eqnarray}
	B' &\approx 7.6\times10^{-6} a_0 S_\alpha R_\lambda \lambda_{\mu\mathrm{m}}^{-1}, &R<r_{\mathrm{mb}} \label{eq:B_R}\\
	&\approx 2.4\times10^{-6} f_i^{1/2} a_0 S_\alpha^{1/2} \lambda_{\mu\mathrm{m}}^{-1}, \hspace{0.5em} &R>r_{\mathrm{mb}} \label{eq:B_rmb}  
\end{eqnarray}
Here and throughout this work, we have extracted the design quantities $a_0$, $S_\alpha$, $R_\lambda = (R/\lambda)$, $\tau_\nu = (c\tau/\lambda)$, and $\lambda_{\mu\mathrm{m}}= \lambda/(1~\mu$m) to highlight the expected scaling with laser properties and channel dimensions.  

The characteristic energy of photon emission is calculated as the first moment of the photon spectrum, $\mathrm{d}N/\mathrm{d}\epsilon_* = (1/\epsilon_*)\mathrm{d}P/\mathrm{d}\epsilon_*$, multiplied by the electron distribution function $f(\epsilon_{\mathrm{e}},t_\nu)$ and averaged over photon energy, electron energy, and time.
Utilizing the numerical result $\int [F(x)/x] \mathrm{d}x\approx3.25$, an otherwise analytical solution is obtained.
Similarly, the total radiated energy is calculated as the integral of Eq.~(\ref{eq:dPde_synch}) multiplied by the electron distribution.
Noting that the number of electrons available for acceleration inside the laser pulse is $N_{\mathrm{e}} = n_{\mathrm{e}} \pi R^2 c\tau$, the expected scaling of $\langle\chi\rangle$, $\langle\epsilon_*\rangle$, $E_{\gamma,\mathrm{tot}}$, and the number of radiated photons $N_\gamma \approx (E_{\gamma,\mathrm{tot}}/\langle\epsilon_*\rangle)$ as a function of the laser--channel interaction parameters is
\begin{eqnarray}
	\langle\chi\rangle &\approx 2.4 C_T a_0 t_\nu  B'  \label{eq:scalings_chi}\\
	\frac{\langle\epsilon_*\rangle}{mc^2} &\approx 0.31 C_T^2 t_\nu^2 a_0^2 B'
	\label{eq:scalings_e}\\
	\frac{E_{\gamma,\mathrm{tot}}}{m c^2} &\approx 2.9\times 10^{13} 
	f_r C_T^2 t_\nu^3 \tau_\nu a_0^3 S_\alpha R_\lambda^2 \lambda_{\mu\mathrm{m}}^2 B'^2
	\label{eq:scalings_E}\\
	N_\gamma &\approx 9.6\times10^{13} f_r t_\nu \tau_\nu a_0 S_\alpha R_\lambda^2 \lambda_{\mu\mathrm{m}}^2 B'.
	\label{eq:scalings_N}	
\end{eqnarray}
Here, $f_r \in [0,1]$ is the radiation duty cycle of electrons, representing the fraction of their orbit spent radiating in high-field regions.
These formulas show that both the scattered photon energy and total radiated power have a strong dependence on laser intensity [note from Eqs.~(\ref{eq:B_R}) and (\ref{eq:B_rmb}) that $B' \propto a_0$] and the duration of the interaction.

The model presented in Eqs.~(\ref{eq:scalings_chi})--(\ref{eq:scalings_N}) assumes continuous linear growth in the electron temperature with time.
This cannot continue indefinitely due to conservation of energy: at some point, depletion of the laser pulse will interfere with continued acceleration and the mechanism will fail.
To determine the length of the interaction, we consider that the total energy in the electron and radiated photon populations must not exceed the initial energy in the laser pulse.
The total energy in the thermal electron distribution is $E_{\mathrm{e,tot}}=N_{\mathrm{e}} T_{\mathrm{e}}$.
Using the same approximations as above and laser energy $E_{\mathrm{L}} \approx 1.54 I \pi R^2 \tau$,\footnote{Assuming Gaussian temporal and spatial profiles and choosing the full-width at half maximum for $\tau$ and half-width at half maximum for $R$, the constant of proportionality is $\pi^{1/2} \ln(2)^{-3/2}/2\approx1.54$.} we find the ratios:
\begin{eqnarray}
	\frac{E_{\mathrm{e,tot}}}{E_{\mathrm{L}}} \approx 1.3 C_T  t_\nu S_\alpha
	\label{eq:Ee_over_EL} \\
	\eta_\gamma \equiv \frac{E_{\gamma,\mathrm{tot}}}{E_{\mathrm{L}}} \approx 1.1\times10^{4} f_r C_T^{2} t_\nu^3 a_0 S_\alpha \lambda_{\mu\mathrm{m}} B'^2,
	\label{eq:Egamma_over_EL} 
\end{eqnarray}
\noindent where $\eta_\gamma$ is defined as the radiation efficiency.
Assuming the electron population dominates the energy budget, Eq.~(\ref{eq:Ee_over_EL}) sets an upper limit on the interaction time of $t_{\nu} < t_{\nu,\mathrm{max}} = 0.768 C_T^{-1} S_\alpha^{-1}$.
The requirement that $S_\alpha > 0.01$ further sets a maximum limit of $t_{\nu,\mathrm{max}} = 77/C_T \approx 960$, which is approximately 3~ps for a 1-$\mu$m laser.
Channel heating, radiation, loss of acceleration potential as the laser is depleted, and other inefficiencies that are not accounted for will further reduce the duration of the interaction, so this may be considered as an upper limit.  
On the further assumption that these other loss mechanisms also scale linearly with time, the cutoff time is selected as $t_\nu = f_t t_{\nu,\mathrm{max}}$ for a fraction $f_t \in [0,1]$.
Incorporating this cutoff time and the magnetic-field model in the limit $R < r_{\mathrm{mb}}$ [Eq.~\ref{eq:B_R}], the analytical scaling laws are derived as:
\begin{eqnarray}
	\langle\chi\rangle &\approx 1.4\times10^{-5} f_t a_0^2  R_\lambda \lambda_{\mu\mathrm{m}}^{-1}	\label{eq:N_scalings_chi}\\
	\frac{\langle\epsilon_*\rangle}{mc^2} &\approx 1.4\times10^{-6} f_t^2 a_0^3   S_\alpha^{-1} R_\lambda \lambda_{\mu\mathrm{m}}^{-1}
	\label{eq:N_scalings_e}\\
	\frac{E_{\gamma,\mathrm{tot}}}{mc^2} &\approx 7.7\times10^2 f_t^3 f_r C_T^{-1} a_0^5 R_\lambda^{4} \tau_\nu
	\label{eq:N_scalings_E}\\
	N_\gamma &\approx 5.6\times10^{8} f_t f_r C_T^{-1}  a_0^2 S_\alpha  R_\lambda^{3} \tau_\nu \lambda_{\mu\mathrm{m}} 
	\label{eq:N_scalings_N}\\
	\eta_\gamma &\approx 2.9\times10^{-7} f_t^3 f_r  C_T^{-1} a_0^3 R_\lambda^2 \lambda_{{\mu}m}^{-1}.
	\label{eq:N_scalings_eff}
\end{eqnarray}
We note that the four dimensionless scaling parameters correspond to those derived in Ref.~\cite{PTRS:Pukhov:2006} for laser--plasma interactions in the ultrarelativistic limit: $a_0$, $S$, $k R$, and $\omega\tau$, respectively.
The energy contained in the laser pulse scales as $E_{\mathrm{L}} \propto a_0^2\tau R^2$, so the emitted energy scales proportionally to $a_0 E_{\mathrm{L}}^2/\tau$, and the number of photons as $E_{\mathrm{L}} R_\lambda$.  

In the limit that $R_\lambda$ exceeds the magnetic boundary [Eq.~(\ref{eq:B_rmb})], we instead find the following
\begin{eqnarray}
	\langle\chi\rangle &\approx 4.5\times10^{-6} f_i^{1/2} f_t a_0^2 S_\alpha^{-1/2} \lambda_{\mu\mathrm{m}}^{-1}	\label{eq:rmb_scalings_chi}\\
	\frac{\langle\epsilon_*\rangle}{mc^2} &\approx 4.4\times10^{-7} f_i^{1/2} f_t^2 a_0^3  S_\alpha^{-3/2} \lambda_{\mu\mathrm{m}}^{-1}
	\label{eq:rmb_scalings_e}\\
	\frac{E_{\gamma,\mathrm{tot}}}{mc^2} &\approx 7.8\times10^1 f_i f_t^3 f_r C_T^{-1} a_0^5 S_\alpha^{-1} R_\lambda^{2} \tau_\nu
	\label{eq:rmb_scalings_E}\\
	N_\gamma &\approx 1.8\times10^{8} f_i^{1/2} f_t f_r C_T^{-1}  a_0^2 S_\alpha^{1/2} R_\lambda^{2} \tau_\nu \lambda_{\mu\mathrm{m}} 
	\label{eq:rmb_scalings_N} \\
	\eta_\gamma &\approx 2.9\times10^{-8} f_i f_t^3 f_r C_T^{-1} a_0^3 S_\alpha^{-1}  \lambda_{{\mu}m}^{-1}.
	\label{eq:rmb_scalings_eff}
\end{eqnarray}
In this case, the number of emitted photons scales proportionally to $E_{\mathrm{L}}$.
In both cases, we find that the efficiency of laser conversion into photons scales proportionally to $a_0^3$, and is otherwise limited by the magnetic boundary.

\subsection{Comparison to 3-D PIC simulations}
\begin{figure}
	\centering
	\includegraphics[trim=163 279 163 277 ,clip,width=\textwidth]{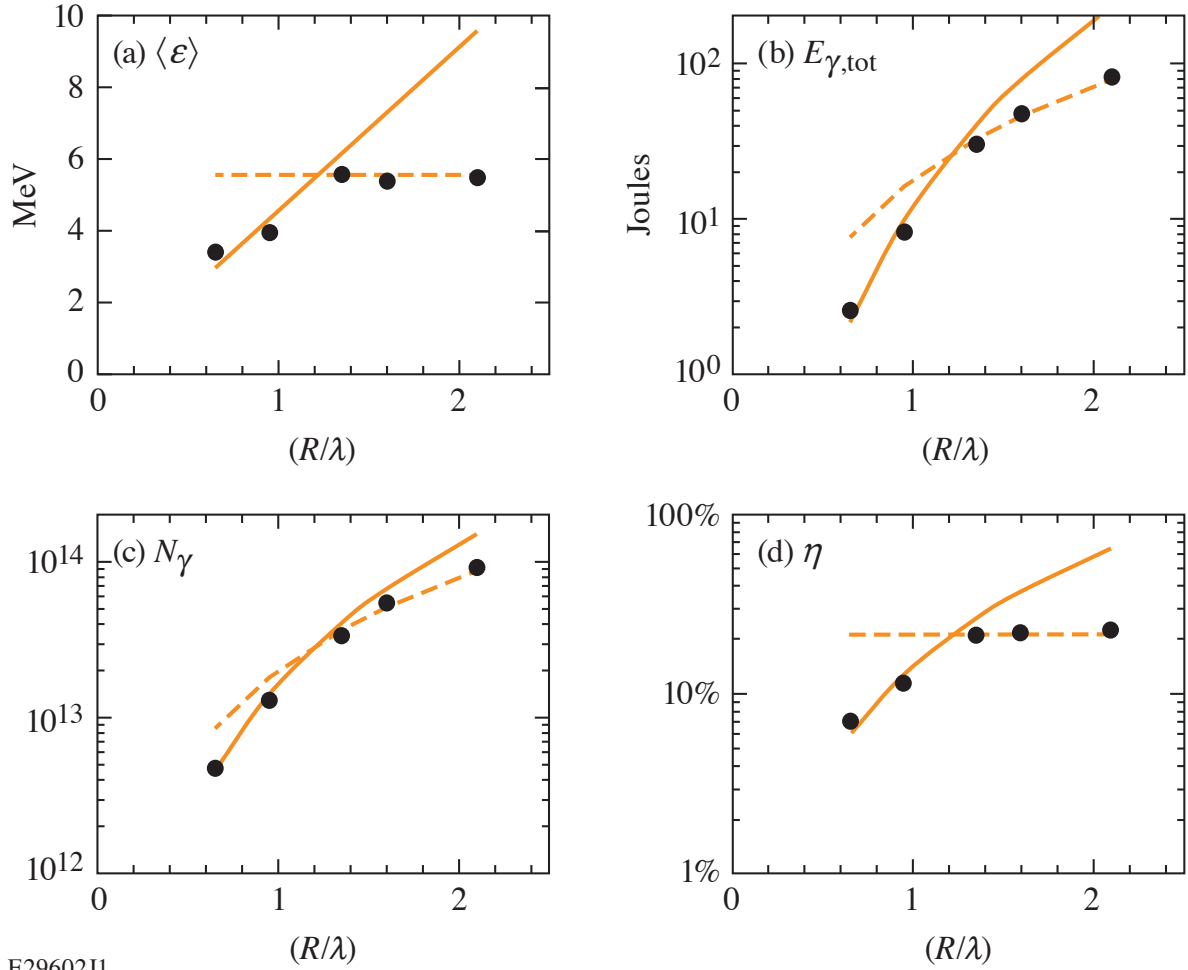}
	\caption{Comparison of derived scaling laws in the limit $R < r_{\mathrm{mb}}$ [Eqs.~(\ref{eq:N_scalings_e})--(\ref{eq:N_scalings_eff}), solid line] and $R > r_{\mathrm{mb}}$ [Eqs.~(\ref{eq:rmb_scalings_e})--(\ref{eq:rmb_scalings_eff}), dashed line] with the results of 3-D PIC simulations described in Ref.~\cite{PRA:Wang:2020} (points): (a) characteristic photon energy, (b) total radiated energy, (c) total number of photons, (d) radiation efficiency.  Simulation results are calculated for photons with energy above 1 MeV in all cases.  Model coefficients are $f_i = 1.53$, $f_t = 0.31$, $f_r = 0.19$. \label{fig:scaling_v_Wang2020}}
\end{figure}

The validity and range of application for these scalings can be evaluated by comparison with the results of full-physics 3-D PIC simulations.
Figure~\ref{fig:scaling_v_Wang2020} presents a comparison with a data set of five simulations using the \textsc{EPOCH} code \cite{PPCF:Arber:2015} that scaled the parameter $R_\lambda$, as presented in Ref.~\cite{PRA:Wang:2020}.
The simulations used design values $a_0 = 190$, $S_\alpha = 0.105$, $\tau_\nu = 10.5$, and varied $R_\lambda$ in the range 0.65 to 2.1, spanning the magnetic boundary value $r_{\mathrm{mb}}/\lambda = \sqrt{f_i/S_\alpha}/\pi \approx 1.2$.
To calculate the characteristic photon energy, total radiated energy, and number of radiated photons using Eqs.~(\ref{eq:N_scalings_e})--(\ref{eq:rmb_scalings_N}) requires the specification of three coefficients: average initial momentum scalar $f_i$, cutoff time fraction $f_t$, and radiation duty cycle $f_{r}$.
An acceleration constant $C_T =  0.08$ is also used, although we note that the scaling laws depend only on $(f_r / C_T)$, rather than on these two coefficients independently.

The model agrees extremely well with the results of the 3-D PIC simulations with $f_i = 1.53$, $f_t = 0.31$~($t=95$~fs), and $f_{r} = 19\%$.
For values $R_\lambda > 1.2$, the trends are observed to follow the prediction limited by the magnetic boundary.  
From the microphysical perspective, the breakdown in scaling with increased $R_\lambda$ arises from effective radial confinement of the electrons, which do not access the higher fields at larger radii\cite{PRA:Wang:2020}.
Since the calculated cutoff time neglects the reduced efficacy of acceleration as the laser is depleted, as well as losses due to ion heating, radiation and other loss mechanisms, the value of $f_t\approx1/3$ is plausible.
A radiation duty cycle of $f_r \approx 1/5$ is similarly reasonable since radiation is produced primarily when the electron is near the region of peak magnetic field.
If an electron's orbit is approximated as a sine wave, this would correspond to the fraction of time in which it experiences a magnetic field with strength greater than 95\% of the maximum value.
The relatively high value of the initial momentum scalar suggests that the electrons that begin with greater momentum may tend to dominate the radiation processes.
Alternatively, this may represent an increase in the magnetic boundary as the electrons gain energy, which is predicted if the phase velocity is not equal to unity.  

This comparison demonstrates that the derived scaling laws are capturing the basic physical processes underlying the magnetic filaments.
Further simulations to study the predicted trends in the other design parameters ($a_0, S_\alpha, \tau_\nu$) will be performed in future work.

\subsection{Interpretation of scaling laws}
The derived scalings imply several consequences for the design of experiments to achieve a given photon source.
Equation~(\ref{eq:N_scalings_e}) predicts that the most-energetic photons for a given intensity will be produced by using channel density near the lower bound $S_\alpha \approx 0.01$ and radius near the magnetic boundary $r_{\mathrm{mb}} \approx 0.4\lambda/\sqrt{S_\alpha} < 3.94\lambda$.
A low channel density allows the interaction to propagate over a greater length before depleting the laser, which in turn allows more acceleration to occur. 
A channel near the upper radius limit will also maximize the total radiated energy.
Comparing Eqs.~(\ref{eq:N_scalings_e}) and (\ref{eq:N_scalings_N}), a denser channel at higher laser intensity could be used to maximize the radiated photons at a given photon energy.

These scalings suggest an experimental program to understand the magnetic filament microphysics and develop the efficient high-energy photon sources they represent.
The primary design variable for scaling the photon energy and radiation efficiency is the laser intensity in the focal spot ($\propto$$a_0^3$).
The characteristic photon energy will achieve gamma-ray scale ($\epsilon_* \sim mc^2$) for experiments with $a_0 > 27$ (intensity of $1\times10^{21}$~W/cm$^2$ with $\lambda_{\mu\mathrm{m}}=1$) and achieve a conversion efficiency of approximately 0.6\%.
If using a 35-fs short-pulse laser, this will produce an estimated $1.9\times10^{12}$ photons.
Experiments at a range of intensity will provide valuable tests of this model: this system at $a_0 = 10$ is predicted to produce $2.6~\times~10^{11}$ photons with characteristic energy of 27~keV and 0.03\% efficiency, while the same system with $a_0 = 50$ is predicted to produce $6.6\times10^{12}$ photons with characteristic energy 3.4~MeV and 4\% efficiency.

From this given initial condition, variations in the target and laser parameters while keeping intensity fixed will provide tests for the underlying physics.
At a given pulse duration, the microchannel length may be varied to artificially truncate the interaction and test the scaling of electron energy, photon energy, and efficiency with interaction time.
If the channels are short enough to prevent depletion, doubling the channel length (and interaction time) is predicted to increase electron energy by a factor of 2$\times$, photon energy by 4$\times$, and total radiated energy and radiation efficiency by 8$\times$.
Such an experiment requires channels shorter than $f_t t_{\nu,\mathrm{max}}\lambda \approx 300~\mu$m at the lower limit of $S_\alpha$.
Deviation from the predicted trends would diagnose the microphysics of laser--channel interaction and electron acceleration in the channel.
Subsequent experiments with longer channels will test how laser depletion limits the electron energy gain and photon emission efficiency.
Varying pulse duration is not predicted to affect depletion or photon energy in this regime, but is predicted to increase the total radiated energy.
Varying the channel density and radius will test the scaling with the design quantities $S_\alpha$ and $R_\lambda$, as well as the predicted behavior as $R_\lambda$ exceeds the magnetic boundary for a given $S_\alpha$
The dependence on wavelength may also be probed by experiments at various laser facilities or by second-harmonic conversion.
Combined, these studies will evaluate and optimize the design of relativistically transparent channel targets at a given laser intensity.

Experiments scaling the laser intensity will test the predicted strong dependence of photon energy, total emitted energy, and efficiency on this parameter.
As derived, the equations neglect the possibility of energy depletion by radiation.
The radiation efficiency formula reaches $\eta_\gamma = 1$ for $a_0 > 679 S_\alpha^{1/3} \approx 146$ for the expected optimal channel configuration.
For experiments approaching this intensity, energy balance will require that laser depletion occur sooner than predicted by these scaling laws.
This will truncate the interaction, presumably resulting in a constant maximum conversion efficiency and radiated energy that scales with laser energy.
We note that the simulations shown in Fig.~\ref{fig:scaling_v_Wang2020} have not yet reached this threshold [$a_0 = 190 < 679 (0.105)^{1/3} \approx 320$] and achieve conversion efficiencies of 20\%.
If this limit were exceeded, the number of radiated photons could then be increased (and the characteristic energy reduced) by increasing the channel density, especially if the extremely high energy photons produced in this regime ($\sim$80~MeV) are not needed.

The scaling of $\langle\chi\rangle$ predicts that the average radiation event will be quantum dominated for $a_0 > 240$, or intensity above $8\times10^{22}$~W/cm$^2$.
In this regime the radiated photons carry a substantial fraction of the radiating electron's energy, which will modify the microphysics of the interacting system and may result in the breakdown of the derived scaling laws.
Several publications have recently begun describing the effects of radiation reaction on direct laser acceleration in this regime\cite{SR:Gong:2019, NJP:Jirka:2020}, which will provide a basis for future extensions of the radiation scaling laws derived here.

The research program described above will require experiments at a wide variety of laser conditions, ranging in pulse duration from 10 to 1000~fs and intensity from $10^{20}$ to $10^{24}$~W/cm$^2$.  
An initial series of experiments was recently performed and will be described in the next section.

\section{Experimental results}
\label{sec:Experiment}
An initial series of experiments was performed using the Texas Petawatt Laser (TPW) to develop an experimental platform to study the relativistically-transparent magnetic filament phenomenon.
In this campaign, microchannels filled with a relativistically transparent density were irradiated using a PW-class laser to demonstrate the predicted direct laser acceleration of electrons and the conversion of laser energy into forward-collimated MeV-class photons.

\begin{figure}
	\centering
	\includegraphics[trim=177 335 177 332,clip,width=\textwidth]{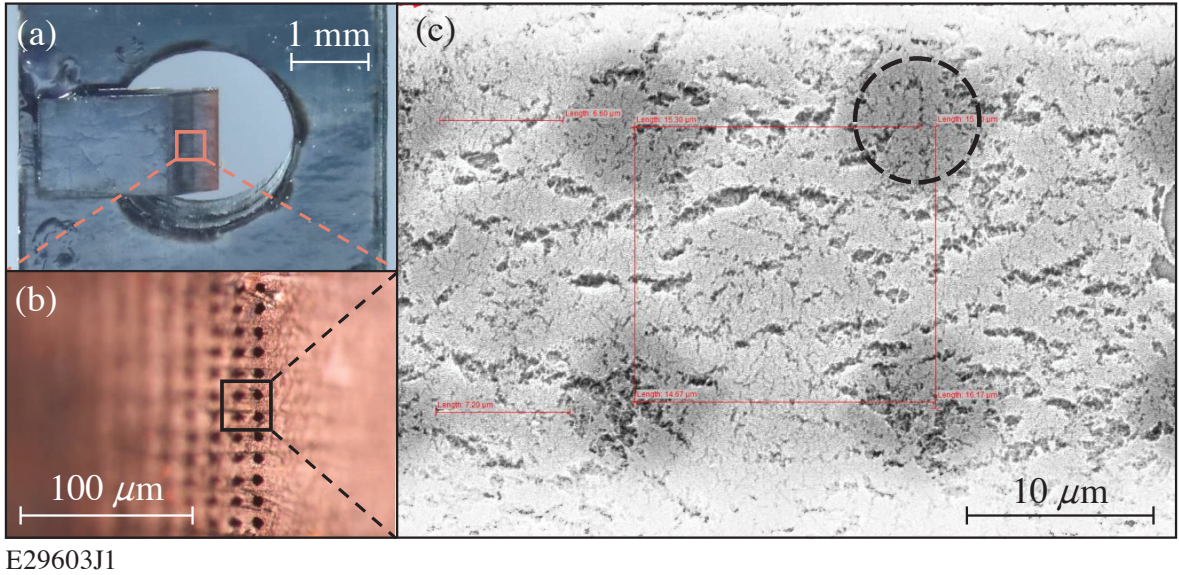}
	\caption{Microchannel array targets used in the TPW campaign: (a) mounting of target on carbon fiber slide; (b) zoom in on front surface of microchannel array; (c) SEM of 6-$\mu$m-ID channels (black dashed, example) filled with low-density CH foam.  The CH foam produces the submicron structure visible on the surface of the target.\label{fig:targets}}
\end{figure}

The targets were microchannel arrays filled with low-density CH foam, as shown in Fig.~\ref{fig:targets}.  
The microchannels were laser-drilled in Kapton, and had inner diameters (ID) of 6~$\mu$m and center-to-center array spacing of 15~$\mu$m in a square lattice.
The front surface of the Kapton substrates were cut with a wedge angle of $20^\circ$ to prevent retro-reflection of the intense laser pulse, resulting in channels that were between 150 and 240~$\mu$m long.
After drilling, the microchannels were filled with CH aerogel solution and supercritically dried to produce a low-density CH foam with mean density of either 15.9 or 31.7~mg/cm$^3$\cite{vendor:GA:aerogel}.  
These correspond to a fully ionized electron density of 5.2 or 10.3~$n_{\mathrm{c}}$, respectively.
Filling of the channels by this procedure was verified using two methods.
Application of an adhesive tape to the surface of a sample after supercritical drying removed 30 to 60~$\mu$m-tall columns of foam, demonstrating that the channel surface was filled to at least this depth.
Additionally, a sample cross-sectioned after filling showed foam fill of the microchannels at all depths that were examined.  
In addition to the microchannel targets, free-standing planar foam targets were produced with a density of 31.7~mg/cm$^3$ and thickness of 124~$\mu$m.
All targets were mounted to carbon fiber slides [shown in Fig.~\ref{fig:targets}(a)] to enable accurate positioning.

\begin{figure}
	\centering
	\includegraphics[trim=140 280 140 280,clip,width=\textwidth]{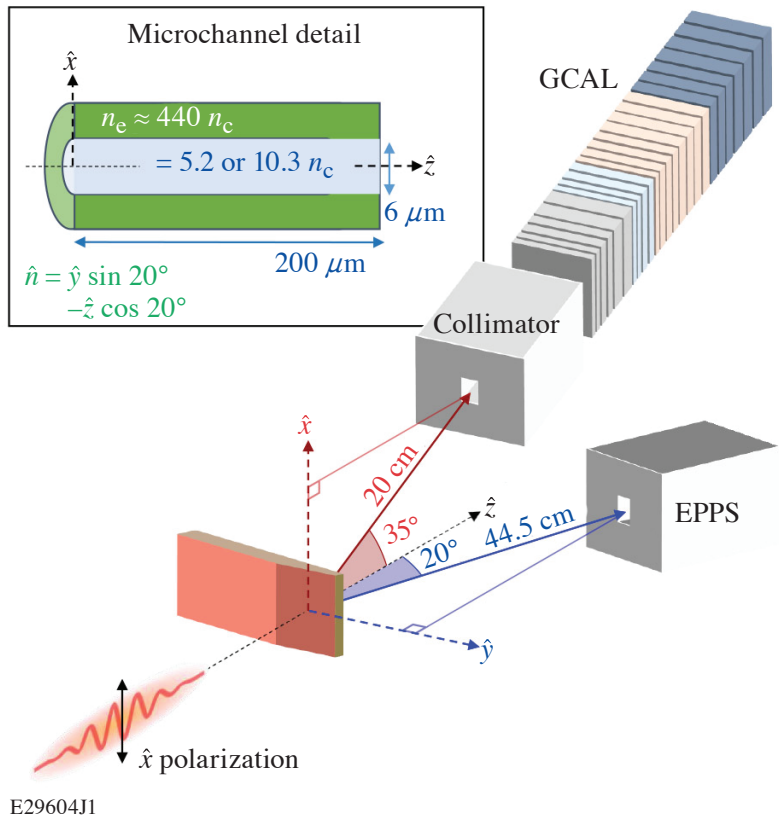}
	\caption{Schematic of the experimental geometry. EPPS is positioned 20$^\circ$ off the laser/microchannel axis in the perpendicular direction ($\hat{y}$); GCAL is positioned 35$^\circ$ off-axis in the polarization direction ($\hat{x}$). (inset) Detail on geometry of microchannels in array target.\label{fig:geometry}}
\end{figure}

The laser was focused using an $f$/3 off-axis parabola with a focal length of 65.5~cm.
The microchannel targets were positioned in the focal plane using translation stages.
The pointing stability of the laser was assessed at 8-$\mu$rad rms, corresponding to 5-$\mu$m rms on target.
An in-chamber camera setup was used to align the laser to the front surface of the targets. 

A schematic drawing of the experimental layout is shown in Fig.~\ref{fig:geometry}.
Electron and photon beams were predicted to be forward-directed, with electron beams concentrated along the axis of the microchannel and photons concentrated in two lobes offset in the laser polarization plane by $\pm20^\circ$.
An electron--proton--positron spectrometer (EPPS) \cite{RSI:Chen:2008} was fielded to record the spectra of accelerated electrons.
The EPPS was fielded at an offset of 20$^\circ$ from the laser axis in the ($y$,$z$) plane, out of the plane of polarization, and a 1-mm-ID front aperture was used.
An image-plate (IP)--based gamma calorimeter (GCAL) designed to record photons in the range 10~keV to 100~MeV was also fielded to record the energy and spectrum of radiated photons.
This instrument records photon signal using 24 samples of BAS-MS IP interspersed between layers of PMMA, aluminum, and stainless steel with thickness varying from 1/16 in. to 5/8 in.
The instrument response is calculated using Geant4 \cite{TNS:Allison:2006} and data are interpreted using an iterative Bayesian unfolding method.\cite{Arxiv:Dagostini:2010}
A second EPPS body was fielded without a front aperture to act as a collimator for the GCAL detector in order to reject electrons and discriminate background from other sources.   
More details on the instrument design and analysis procedure will be described in a forthcoming manuscript\cite{Arxiv:LasoGarcia:upcoming}.

In this campaign, 11 shots were performed for which the target was accurately positioned at the laser focus.
These shots delivered an average of 98.8$\pm$6.0~J of $\lambda = 1057$-nm laser light, with a peak power of 694$\pm$38~TW and intensity of (1.09$\pm$0.07)~$\times$~10$^{21}$~W/cm$^2$ on target.  
This corresponds to $a_0 = 29.9$$\pm$$1.0$ and a relativistic transparency parameter $S_\alpha = 0.173$ (0.345) for the $n_{\mathrm{e}} = 5~(10)~n_{\mathrm{c}}$ targets, respectively.
In the limit of relativistic currents, these values substantially surpass the requirement $S_\alpha > 0.01$.
The radius of the focal spot at 50\% peak intensity was $2.57$$\pm$$0.12~\mu$m.
Given the relative target--laser alignment precision as compared to the channel size,  not all experiments with channel targets were expected to successfully inject laser energy into the channels.

\begin{figure}
	\centering
	\includegraphics[trim=93 318 90 314,clip,width=\textwidth]{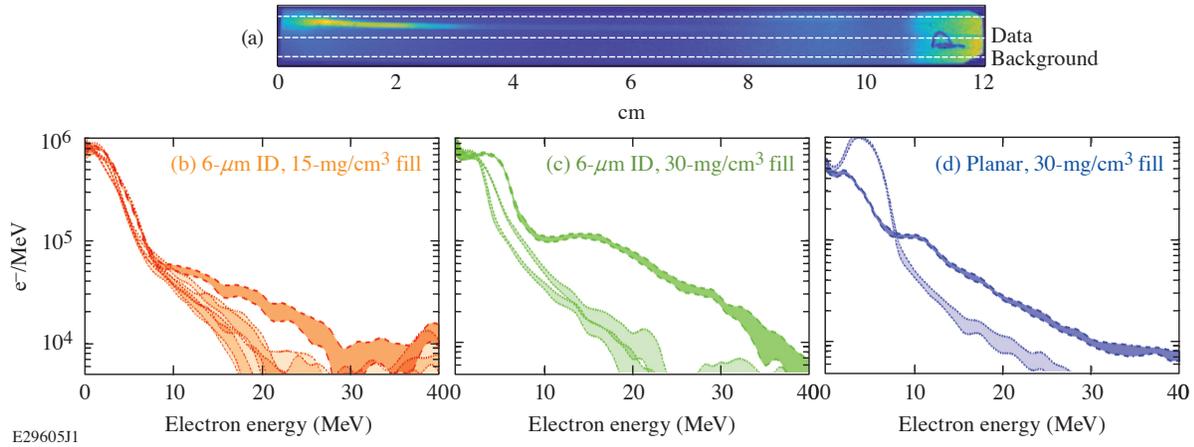}
	\caption{(a) Raw electron spectral data recorded on an image plate using the EPPS. \mbox{[(b)--(d)]}~Electron spectra (1$\sigma$ envelopes) interpreted for 6-{$\mu$m} channels filled with (b) 15-mg/cm$^3$ or (c) 30-mg/cm$^3$ CH foam; (d) planar 30-mg/cm$^3$ foam targets.  Shots in each category with the highest hot electron temperature $T_2$ are emphasized.\label{fig:EPPS_data}}
\end{figure}

\begin{figure}
	\centering
	\includegraphics[trim=91 307 91 304,clip,width=\textwidth]{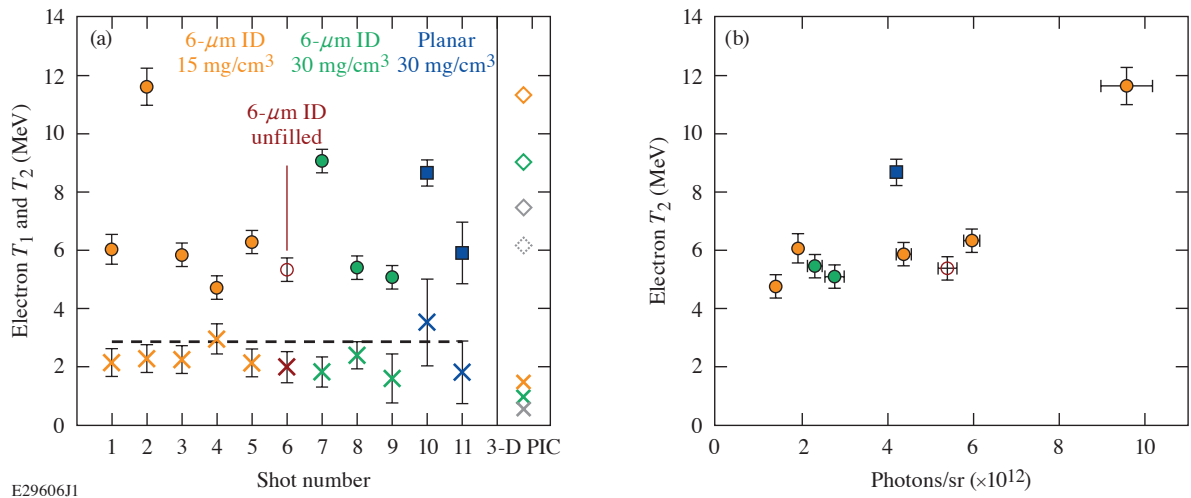}
	\caption{(a) Results of two-temperature fits to EPPS data ($\times$ represents $T_1$): 6-$\mu$m-ID channels filled with (orange) 15-mg/cm$^3$ foam, (green)  30-mg/cm$^3$ foam; (red open) empty channels; (blue) planar 30-mg/cm$^3$ foams.  (black dashed line) Electron temperature predicted in Ref.~\cite{PRL:Haines:2009}.  Right column shows  ($\times$) $T_1$ and ($\diamond$) $T_2$ fits to electron spectra from 3-D PIC simulations: (orange) 5~$n_{\mathrm{c}}$ channel, (green) 10~$n_{\mathrm{c}}$ channel, (grey) 100~$n_{\mathrm{c}}$ planar, (grey dotted) 200~$n_{\mathrm{c}}$ planar.  (b) Comparison of hot-electron temperature ($T_2$) with the photon brightness above 10~keV recorded by GCAL.\label{fig:EPPS_GCAL_combo}}
\end{figure}

The electron spectra recorded on these experiments are shown in Fig.~\ref{fig:EPPS_data}.
The raw electron data was recorded on IP's and interpreted using the fade-corrected photostimulated luminescence\cite{RSI:Tanaka:2005}, measured IP response to electrons\cite{RSI:Bonnet:2013}, and calculated dispersion of the spectrometer.
Electron spectra were fit using a two-temperature distribution: results are shown in Fig.~\ref{fig:EPPS_GCAL_combo}(a).
The bulk electron temperatures ($T_1$) were observed to be consistent across the microchannel targets at $2.3\pm0.5$~MeV.  
This value is roughly consistent with the relativistic electron scaling in Ref.~\cite{PRL:Haines:2009}, which predicts a temperature of $2.8$~MeV for laser interaction with an overdense plasma slab at these intensities.  
A hot-electron temperature ($T_2$) was observed in all experiments with an average energy of $6.7$$\pm$$2.1$~MeV, or roughly three times $T_1$.
The hot-electron temperature was observed to be unusually high in two of eight foam-filled microchannel targets: shots \#2 (15-mg/cm$^3$ fill) and \#7 (30-mg/cm$^3$) recorded $T_2$ of 11.6 and 9.1~MeV, respectively, as compared to 4.7 to 6.3~MeV for the remainder of the experiments.  
These values are also increased relative to the ``cold''-electron temperature, with $T_2/T_1 \approx 5$ for both of these cases, as compared to $2.5$$\pm$$0.6$ for the others.
This difference is more than 5$\times$ the standard deviation observed in the set of typical shots.
An unusually high value of $T_2 = 8.7$~MeV was also observed in one of the planar foam samples (\#10).  

The shot with the highest recorded electron temperature (\#2) was a 6-$\mu$m-ID microchannel filled with 15-mg/cm$^3$ foam.
This shot also produced the largest photon signal in GCAL.
Comparing results from the EPPS and GCAL shows a positive correlation between the observed temperature of the electron distribution (4 to 12~MeV) and the number of MeV photons, as shown in Figure~\ref{fig:EPPS_GCAL_combo}(b).
(The GCAL data from shots \#7 and \#11 were lost due to experimental error.)
This correlation is observed across all target types, including foam-filled channels, unfilled channels, solid targets, and planar foam targets.

We conclude that the predicted electron acceleration and photon radiation dynamics of the relativistically transparent laser--channel interaction were observed in two of the foam-filled microchannel shots. 
Given the limited number of shots in this campaign and the stochastic nature of the laser pointing on the scale of the array structure, it is necessary to ask what is the likelihood that the laser would successfully inject intense laser light into the channel in two out of eight experiments.
(We exclude the unfilled channel from this calculation due to an insufficient amount of experimental data for this case.)
As a worst-case scenario, the center of the laser focus may strike the array at any point, and the probability that it falls within a microchannel entrance is given by the ratio of the channel area to an array unit cell: for channel radius $R = 3~\mu$m and center-to-center channel separation $L = 15~\mu$m, the probability is $P_i = \pi R^2/L^2 \approx 0.126$.
Over $N = 8$ shots, the probability that this never happens is $P(0) = (1-P_i)^N \approx 0.34$, implying a nearly 2-in-3 chance that at least one interaction occurs.
The probability of one, two, and three interactions are $P(N=[1,2,3]) \approx [0.39, 0.20, 0.06]$, respectively, with approximately 1\% chance of four or more interactions in eight samples.
Using the as-measured pointing stability (5-$\mu$m rms), the likelihood of interaction is somewhat higher: a numerical study found the probability of interaction when taking the measured pointing stability into account to be approximately $P([0,1,2,3]) \approx [0.21, 0.36, 0.27, 0.12]$, with the likelihood of achieving at least one interaction nearly four in five.
From these calculations, the proposed explanation of the observed data set is statistically plausible.

The planar 30-mg/cc foam experiments (\#10 and \#11) agree with the range of values observed in the microchannel experiments with comparable fill density.
Although the two planar foam experiments are nominally identical, variation in the electron spectra is expected due to the inherent stochasticity of relativistically transparent interactions with a high value of $S_\alpha$.  
Previous planar experiments with $S_\alpha>0.14$ have shown that the hosing instability results in apparently random variations in the direction of accelerated electrons and radiated photons\cite{NJP:Willingale:2018}.
Because the electron and photon spectra in the present experiment are only collected over a small solid angle in a fixed direction, we would expect to observe variation between nominally identical shots with planar foam targets. 
This may also explain the comparatively low photon signal for this target in Fig.~\ref{fig:EPPS_GCAL_combo}.
In future experiments, electron and x-ray spectrometers that cover a large solid angle with respect to the target will be required to diagnose more fully the radiation characteristics of these targets.

\subsection{Post-shot simulations}

\begin{figure}
	\centering
	\includegraphics[trim=91 325 90 323,clip,width=\textwidth]{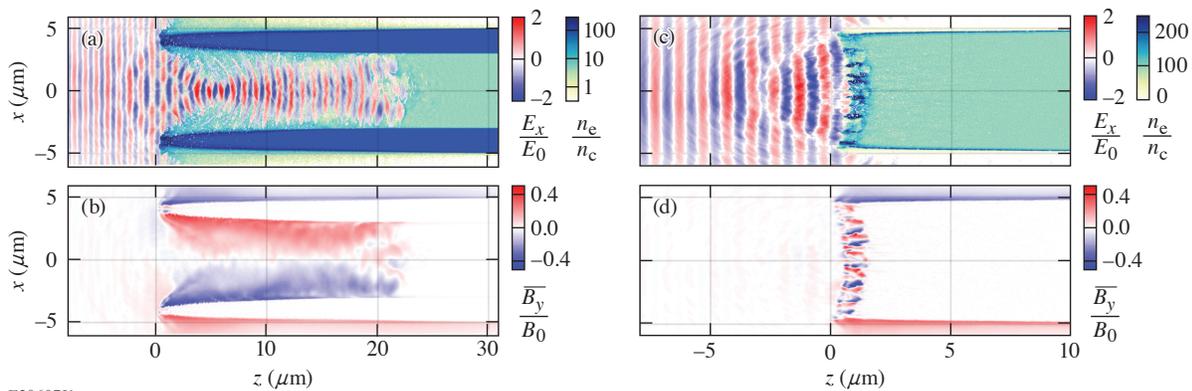}
	\caption{Slices in the ($x$,$z$) plane from 3-D PIC simulations of the experimental laser condition irradiating [(a),(b)] 6-$\mu$m-ID, 5~$n_{\mathrm{c}}$ foam-filled microchannels and [(c),(d)] 100~$n_{\mathrm{c}}$ relativistically overdense slab: [(a),(c)] electric field normalized to peak laser field; electron density normalized to critical density; [(b),(d)] azimuthal magnetic field normalized to peak laser field.\label{fig:PICmap}}
\end{figure}

To compare with the experimental results, 3-D PIC simulations using the experimental laser conditions were performed for the foam-filled channels and a relativistically overdense (100~$n_{\mathrm{c}}$) ``solid'' target.
The simulations were performed using the code \textsc{EPOCH} with the same methodology as in Sec.~\ref{sec:Scaling} and previous work\cite{PRL:Stark:2016,PRE:Gong:2020,PRE:Arefiev:2020,PRA:Wang:2020}.
A plot of the electric field, electron density, and azimuthal magnetic field in the cross-section plane is shown in Fig.~\ref{fig:PICmap} for the $5~n_{\mathrm{c}}$ (15-mg/cm$^3$) filled channel and for the overdense planar target.
The images are recorded 57~fs after peak intensity reaches the target front surface for the channel target, and 7~fs after peak intensity for the overdense target.
The relativistically transparent magnetic filament is observed in the microchannel case, with peak azimuthal field strength of approximately 40\% the peak laser field strength.

\begin{figure}
	\centering
	\includegraphics[trim=94 335 92 333,clip,width=\textwidth]{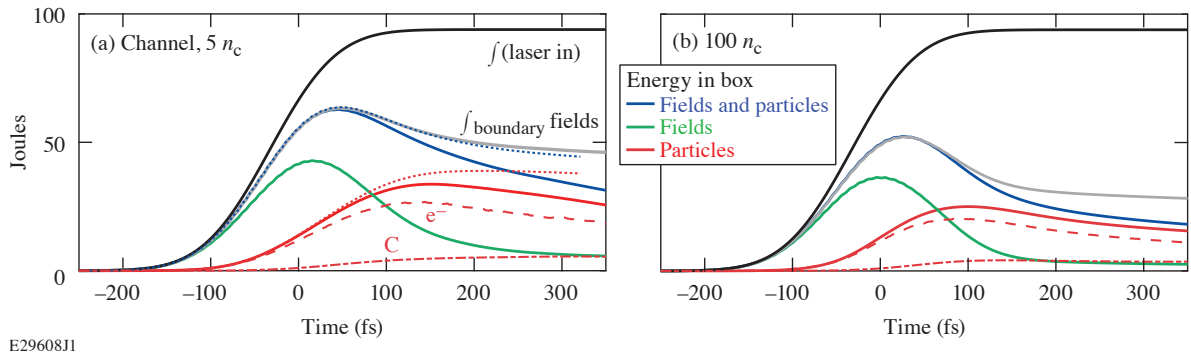}
	\caption{Energy history versus time for the 3-D PIC simulations of experiments (a) 6-$\mu$m channel with 5~$n_{\mathrm{c}}$ fill and (b) overdense target: black curve, total injected laser energy; grey curve, integrated field flux at box boundary; blue curve, total energy in simulation box; green curve, field energy in box; red curves, particle energy in box, including: dashed curve, electrons; dashed--dotted curve, carbon; (not shown) hydrogen and x-rays.  Total energy in box (dotted blue curve) and total particle energy in box (dotted red curve) for simulation with reflective boundaries, preventing particles from escaping. \label{fig:simHistory}}
\end{figure}

The difference in energetics between the channel and solid cases is shown in Fig.~\ref{fig:simHistory}.
The time scale in these plots is relative to the time at which the peak laser intensity arrives at the surface of the channel or solid target.
From the integrated field flux at the boundary (grey curve) we can infer that the microchannel reflects roughly half of the incident energy, whereas the solid density case reflects approximately 75\%.  
In the channel case, the heating of the electrons proceeds from roughly --100 to +150~fs, with the most rapid heating around 25~fs.
In both cases, most of the energy remaining in the box by the end of the interaction is contained in the electron population.
Comparing to a channel simulation with reflective boundaries for particles (dotted curves) shows that the loss of energy after $\sim$100~fs is increasingly due to energetic electrons escaping the simulation.

The spectra of electrons escaping the simulation were sampled in the direction of the electron spectrometer in the experiment, and were fit with the same analysis method used to process the experimental data.
The resulting simulated electron temperatures $T_1$ and $T_2$ are plotted on the right side of Fig.~\ref{fig:EPPS_GCAL_combo}(a). 
The simulated values of $T_2$ for the 5~$n_{\mathrm{c}}$ and 10~$n_{\mathrm{c}}$ cases  (11.3~MeV, 9.0~MeV) agree well with the measured values from the relevant shots with heightened hot electron temperatures (\#2 and \#7, respectively), supporting the interpretation that the laser was well-aligned to the channels in this subset.
For the 100~$n_{\mathrm{c}}$ case, the simulated value of $T_2= 7.5$~MeV is 1.9~MeV hotter than the baseline shot average.  
This discrepancy appears to be due to the fact that the Kapton targets have a fully-ionized electron density of approximately $440~n_{\mathrm{c}}$, or 4.4$\times$ the density used in the simulations.  
To test this hypothesis, a 200~$n_{\mathrm{c}}$ simulation was performed with a truncated simulation domain.
This simulation produced a $T_2$ electron temperature of 6.2~MeV (dotted diamond in Fig.~\ref{fig:EPPS_GCAL_combo}(a), only 0.6~MeV hotter than the baseline shot average.
Extrapolating this trend, a simulation at the full experimental target density (which could not be performed due to computational cost) would agree with the lower baseline experimental values, supporting the hypothesis that these cases are explained by the laser striking the solid surface between channels.
In the channel simulations, the electron acceleration occurs in the relativistically-transparent region near peak laser intensity, and is not affected by the density of the channel walls as long as they are relativistically over-dense.

\begin{figure}
	\centering
	\includegraphics[trim=91 335 92 335,clip,width=\textwidth]{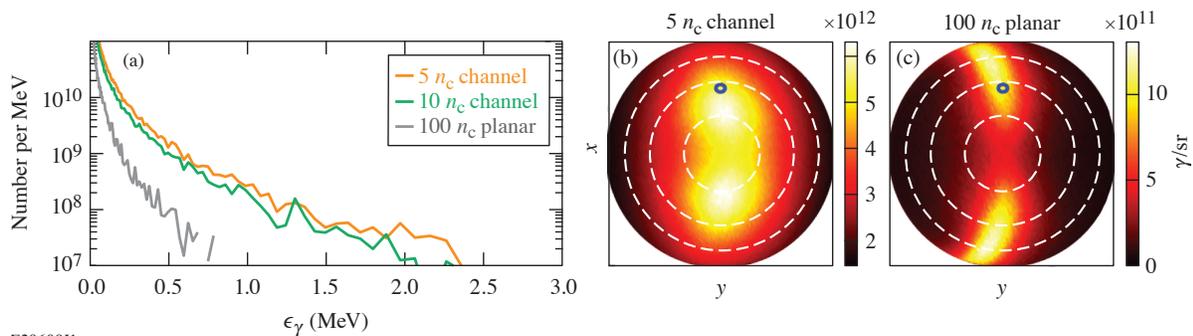}
	\caption{(a) Spectrum of photons radiated in the direction of the GCAL photon spectrometer in PIC simulations: 6-$\mu$m-ID channels filled with 5~$n_{\mathrm{c}}$ (orange curve), 10~$n_{\mathrm{c}}$ density plasma (green curve); and 100~$n_{\mathrm{c}}$ planar plasma  (blue curve). (b,c) Angular distribution of radiated photons with energy above 10~keV for the (b) 5~$n_{\mathrm{c}}$-filled channel, (c) 100~$n_{\mathrm{c}}$ planar target; note that the color scales differ by a factor of $\sim$5. Direction of GCAL in the experiments (blue circle).  Angles of $20^\circ, 40^\circ, 60^\circ$ from the laser axis (white dashed). \label{fig:simPhotons}}
\end{figure}

The spectral and spatial distribution of the radiated x rays produced by relativistic electrons is shown in Fig.~\ref{fig:simPhotons}.
The spectra were sampled from a region with 2$^\circ$ opening angle along the vector sampled by the GCAL photon spectrometer in the experiment [shown by the blue circle in Figs.~\ref{fig:simPhotons}(b) and \ref{fig:simPhotons}(c)].
This region exceeds the solid angle of the GCAL detector by a factor of $\sim$4.  
However, the scale is small compared to the gradient scale length of the photon spatial distribution so this difference does not affect the observed spectral shape.
The characteristic photon energy increases from 44~keV for the solid-density foil to 76~keV for the microchannel targets.
The number of radiated photons observed also increases by a factor of $\sim$5.
This matches the difference seen in Fig.~\ref{fig:EPPS_GCAL_combo}(b) between the hottest and brightest shot [\#2 in Fig.~\ref{fig:EPPS_GCAL_combo}(a)] as compared to the coldest and dimmest shot (\#4).
Spatially, the distribution of the x rays is simulated to produce two lobes separated along the polarization axis in all cases.
These lobes represent the average direction of the relativistic electrons when radiating, and are concentrated closer to the experiment axis ($\pm$22$^\circ$) for the 5~$n_{\mathrm{c}}$ channel as compared to the planar target ($\pm$54$^\circ$).  
In the channel target, the electron net acceleration in the forward direction causes this increased concentration as well as the increased brightness.
The spatial pattern of x-ray radiation is slightly curved toward the $-y$ direction in both cases due to the 20$^\circ$ angle of the target surface normal with respect to the channel axis.

The simulation study supports the interpretation that the experimental data includes two shots with full laser--channel interaction and several shots with limited or no laser--channel interaction.
The ratio of photon brightness between the channel and solid cases matches the ratio between the brightest and dimmest experiments observed.
It is possible that some of the data points shown in Fig.~\ref{fig:EPPS_GCAL_combo} represent intermediate cases, in which a fraction of the laser energy entered channels and produced an intermediate electron and photon spectrum.

\section{Discussion}
\label{sec:Discussion}

Applying the scaling laws developed in Sec.~\ref{sec:Scaling} to the experiments demonstrates that the design parameters were not optimal to maximize either the  characteristic x-ray energy or total radiated energy.
The values of $S_\alpha$ used were much higher than required: 0.174 and 0.344 for the 5~ and 10~$n_{\mathrm{c}}$-filled channels, respectively.
This more rapidly depleted the laser in the channel (cutoff time of 61~fs and 31~fs, respectively) limiting the electron acceleration.
Additionally, this set a very narrow magnetic boundary (1.00 and 0.71~$\mu$m, respectively), further limiting photon energy and total radiated energy.
A lower-density fill with $n_{\mathrm{e}} > 0.3 n_{\mathrm{c}}$ could have decreased $S_\alpha$ to near the lower limit of 0.01, keeping all other properties constant.
This would have extended the cutoff time by more than a factor of 10 and increased the magnetic boundary to greater than the laser focal radius.
The scaling laws predict an increase in the characteristic photon energy to 0.42~MeV, an increase in the total radiated energy by a factor of $\sim$7, and increased efficiency to 0.3\%.
Such a low density ($\sim$1~mg/cm$^3$) has been achieved using silica aerogel\cite{vendor:GA:aerogel}.
Alternatively, gaseous fill may be used for future experiments in this regime: 0.9~atm of diatomic nitrogen gas would provide the required electron density when fully ionized.

For future experiments, the stochastic nature of laser--channel pointing is undesirable because it reduces the physics throughput and limits applications of this platform as a photon source.
Improved channel designs have been developed using two-photon polymerization to print the target structures with increased channel density and reduced wall thickness between channels.
A close-packed hexagonal structure of 6-$\mu$m-ID channels with 8-$\mu$m center-to-center separation (2-$\mu$m walls) has been demonstrated, which increases the fraction of the surface covered by channel openings to over 50\%. 
Larger laser spots may be used intentionally to improve the repeatability of the platform by including radiation from multiple neighboring channels in each experiment.
For laser pulses longer than 100~fs, simulations show that the wall can significantly distort on time scales of the interaction.
The optimization of the channel thickness as a function of material and laser pulse duration will be the subject of future research.

For future experiments at substantially higher laser power and intensity, we note that the scaling laws derived in Sec.~\ref{sec:Scaling} are valid only for classical radiation ($\chi < 1$).
In the quantum radiation regime ($\chi > 1$), the radiated power scales as $P \propto \chi$, and radiation reaction will significantly modify the electron orbits as single photons carry a substantial fraction of the electron's energy ($\epsilon_* \sim mc^2$).
From Eq.~(\ref{eq:N_scalings_chi}), the radiation is expected to be dominated by quantum interactions for $a_0 > 240$.
Simulations predict that this effect may benefit electron acceleration by stochastically preventing electron-laser dephasing over the millimeter- and picosecond-scale interactions  that become accessible at such high intensities\cite{SR:Gong:2019}.
Future theoretical and simulation work will evaluate how this affects the net acceleration and radiation properties of the magnetic filaments in the ultra-relativistic regime.

\section{Conclusions}
\label{sec:Conclusions}

We have derived scaling laws to describe the radiative properties of the relativistically transparent magnetic filament phenomenon, including characteristic photon energy, total radiated energy, number of radiated photons, and radiation efficiency in terms of several design parameters: intensity $a_0$, the relativistic transparency parameter $S_\alpha$, normalized focal radius $R_\lambda$, and normalized pulse duration $\tau_\nu$.
The scaling laws were compared to 3-D PIC simulations that varied $R_\lambda$, and were shown to agree well for reasonable choices of constant parameters.
The results of experiments to test this phenomenon in the moderate-intensity regime ($a_0\approx30$) were presented.
Electron spectrometer and photon calorimeter data showed increased electron temperature and photon production by up to a factor of $\sim$5 for a subset of the foam-filled microchannel experiments.
This was found to be consistent with the statistical likelihood of laser--channel interaction, given the measured pointing stability and the geometry of the microchannel array targets.

The scaling laws derived here will guide future experimental and simulation research into relativistically transparent laser--plasma interactions.
This theoretical framework will significantly improve the design of experimental campaigns by reducing the dependence on 3-D PIC simulations, which can be time-consuming and costly.
Future experiments and simulations will continue to test the predicted scaling of radiative properties with laser and channel designs, leading to increased efficiency and control of electron acceleration and photon radiation at higher intensities.
The 10~PW-class lasers that are now becoming available at ELI-NP \cite{RRP:Ursescu:2016} and ELI-Beamlines L4-ATON \cite{MRE:Weber:2017} offer exciting opportunities for ultra-high flux gamma-ray sources from this source.
With $5\times10^{22}$~W/cm$^2$ intensity, the scaling laws predict that magnetic filament experiments on ELI-NP will produce 10$^{13}$, 68~MeV photons with efficiency of $\sim$48\%, while experiments on L4-ATON will produce $5\times10^{13}$, 96~MeV photons with efficiency of $\sim$53\%, with both cases limited by depletion.
These sources will be world-leading in terms of brightness and flux of MeV photons.
Ultimately, this research program promises to create a laboratory platform for strong magnetic-field physics in the megaTesla range, as well as an efficient, bright, and repeatable source of MeV-scale photons for use in a wide range of laser-driven strong-field QED studies.

\ack
This material is based upon work supported by the Department of Energy National Nuclear Security Administration under Award Number DE-NA0003856, the University of Rochester, and the New York State Energy Research and Development Authority. 
This work was supported by DOE Office of Science, Fusion Energy Sciences under Contract No. DE-SC0019167 the LaserNetUS initiative at the Texas Petawatt Laser facility.
T.W. and A.A. were supported by AFOSR (Grant No. FA9550-17-1-0382).
Simulations were performed with \textsc{EPOCH} (developed under UK EPSRC Grants EP/G054950/1, EP/G056803/1, EP/G055165/1 and EP/ M022463/1) using high performance computing
resources provided by TACC at the University of Texas. 
Special thanks to Hui Chen, Mike Donovan, and Michael Spinks.

This report was prepared as an account of work sponsored by an agency of the U.S. Government. Neither the U.S. Government nor any agency thereof, nor any of their employees, makes any warranty, express or implied, or assumes any legal liability or responsibility for the accuracy, completeness, or usefulness of any information, apparatus, product, or process disclosed, or represents that its use would not infringe privately owned rights. Reference herein to any specific commercial product, process, or service by trade name, trademark, manufacturer, or otherwise does not necessarily constitute or imply its endorsement, recommendation, or favoring by the U.S. Government or any agency thereof. The views and opinions of authors expressed herein do not necessarily state or reflect those of the U.S. Government or any agency thereof.


\end{document}